\pdfoutput=1
\documentclass[11pt,twoside,a4paper,cmspaper,final,collab]{cms-tdr}
\def\svnVersion{c639513-D}\def\svnDate{2021/04/07}\def\cmsCernNoTag{CERN-EP-2020-039}\def\cmsCernDate{\today}\def\cmsMessage{"Published in Physics Letters B as \href{http://dx.doi.org/10.1016/j.physletb.2021.136385}{\doi{10.1016/j.physletb.2021.136385}.}"}
\begin{document}\cmsNoteHeader{HIN-19-002}

\newcommand{\sqrts}{\ensuremath{\sqrt{s}}\xspace}
\newcommand{\vtwo}{\ensuremath{v_{2}}\xspace}
\newcommand{\PbPb}{\ensuremath{\text{PbPb}}\xspace}
\newcommand{\NSigone}{\ensuremath{N_{\mathrm{sig,1S}}}\xspace}
\newcommand{\NSigtwo}{\ensuremath{N_{\mathrm{sig,2S}}}\xspace}
\newcommand{\NSigthree}{\ensuremath{N_{\mathrm{sig,3S}}}\xspace}
\newcommand{\dimu}{\ensuremath{\text{dimuon}}\xspace}

 \ifthenelse{\boolean{cms@external}}{\providecommand{\cmsLeft}{upper\xspace}}{\providecommand{\cmsLeft}{left\xspace}}
\ifthenelse{\boolean{cms@external}}{\providecommand{\cmsRight}{lower\xspace}}{\providecommand{\cmsRight}{right\xspace}}

\cmsNoteHeader{HIN-19-002}
\title{Measurement of the azimuthal anisotropy of $\PgUa$ and $\PgUb$ mesons in PbPb collisions at $\sqrtsNN = 5.02\TeV$}

\author*[inst1]{Yongsun Kim}

\date{\today}

\abstract{
The second-order Fourier coefficients (\vtwo) characterizing the azimuthal distributions of \PgUa and \PgUb mesons produced in PbPb collisions at \sqrtsNN = 5.02\TeV are studied. The \PgU mesons are reconstructed in their \dimu decay channel, as measured by the CMS detector.  The collected data set corresponds to an integrated luminosity of 1.7\nbinv. The scalar product method is used to extract the \vtwo coefficients of the azimuthal distributions. Results are reported for the rapidity range  $\abs{y}<2.4$, in the transverse momentum interval $0 < \pt < 50\GeVc$, and in three centrality ranges of 10--30\%, 30--50\% and 50--90\%. In contrast to the \PJGy mesons, the measured \vtwo values for the \PgU mesons are found to be consistent with zero. 
}

\hypersetup{
pdfauthor={CMS Collaboration},%
pdftitle={Measurement of the anisotropic flow of Upsilon(1S) and Upsilon(2S) mesons in PbPb collisions at sqrt(s) = 5.02 TeV},%
pdfsubject={CMS},
pdfkeywords={CMS, physics, heavy ion, upsilon, v2, flow, PbPb, 5.02TeV}}

\maketitle 

\section{Introduction}
\label{sec:intro}

The primary motivation for studying high-energy heavy ion collisions is to better understand the hot and dense matter produced in these interactions~\cite{Wit:2018HI}.
Lattice quantum chromodynamics (LQCD) calculations indicate that at sufficiently high temperatures an environment is created for a crossover to occur from hadronic matter to a strongly interacting system of deconfined quarks and gluons~\cite{Karsch:2000ps}. This deconfined medium is called the ``quark-gluon plasma" (QGP)~\cite{Shuryak:1977ut}.
One of the most prominent signatures of QGP formation is that the production of quarkonia, the bound states of a heavy quark and its antiquark, is suppressed with respect to expectations from scaling the yields in proton-proton collisions by the number of binary nucleon-nucleon (NN) collisions. This suppression arises because the quarkonia binding is weakened by color screening caused by the surrounding partons in the medium~\cite{Matsui:1986dk}, and therefore the extent of the quarkonia suppression is expected to be sequentially ordered by the binding energies of the quarkonia states. 
In heavy ion collisions, the majority of heavy quarks are produced from hard scattering processes at an early stage and therefore can interact with the QGP throughout its entire evolution.  

Because of the binding energy dependence of the screening, the bottomonium states ($\PgUa$, $\PgUb$, $\PgUc$, \PGcb, etc.) are particularly useful probes to understand the space-time evolution of the QGP.
The sequential suppression of the yield of \PGUP{nS} states was first observed by CMS~\cite{Chatrchyan:2012lxa, Khachatryan:2016xxp} at a NN center-of-mass energy $\sqrtsNN = 2.76\TeV$. 
More recently, results with improved statistical precision have been reported by both the ALICE Collaboration~\cite{Acharya:2018mni} and CMS Collaboration~\cite{Sirunyan:2017lzi, Sirunyan:2018nsz} at $\sqrtsNN = 5.02\TeV$.  
The suppression of the \PgUa meson has also been studied at $\sqrtsNN = 200\GeV$ at RHIC~\cite{Adamczyk:2013poh}.
The available experimental data, spanning from 0.20 to 5.02\TeV, have provided new insight into the thermal properties of the QGP~\cite{HeavyFlavorQuarkoLHC}.

The screening due to the QGP can also result in an azimuthal asymmetry in the observed yields of quarkonia.
In non-central heavy ion collisions, the produced QGP has a lenticular shape in the transverse plane.
Consequently, the average path length for quarkonia traveling through the medium depends on the direction taken with respect to this shape, with a larger suppression in the direction of the longer axis~\cite{He:2015hfa, Borghini:2010hy, Romatschke:2015dha, Jaiswal:2017dxp}. 
The anisotropic distribution of particles can be characterized by the magnitudes of the Fourier coefficients ($v_{n}$) of the azimuthal correlation of particles, through the relation
\begin{linenomath}
\begin{equation}
\label{eq:v2}
{\frac{ \rd N } { \rd\phi } } \propto 1 + 2\sum_{n}{v_n} \cos[n( \phi - \Psi_{n})].
\end{equation}
\end{linenomath}
\noindent Here $\phi$ is the angle of the particles and $\Psi_{n}$ is the angle characterizing the symmetry plane of the $n$th harmonic based on the charged particle density distribution~\cite{Voloshin:1994mz}.
By studying the azimuthal distribution of the quarkonia, it is possible to develop a more comprehensive understanding of the dynamics of their production.

Measurements of \vtwo values have been reported for \PJGy mesons by the ALICE Collaboration~\cite{ALICEJpsi276TeVPbPb,ALICEJpsi502TeVPbPbFullRun2} and the CMS Collaboration~\cite{CMSJpsi276TeVPbPb} at $\sqrtsNN = 5.02\TeV$ and 2.76\TeV, respectively. Significant evidence for an azimuthal asymmetry in the particle yields is found for both experiments. 
However, the \PJGy asymmetry can also arise from the late-stage recombination of uncorrelated charm quark pairs inside the QGP,
which acts in the opposite direction as color screening. 
The relative contribution of recombination to the inclusive cross section of bottomonium states is expected to be smaller than for charmonium states~\cite{Krouppa:2016jcl, Brambilla:2016wgg, Du:2017qkv,Emerick:2011xu}.
For this reason, the azimuthal anisotropy of bottomonia better reflects the screening that occurs during its passage through the QGP medium.
The ALICE Collaboration has measured the \vtwo\ for \PgUa mesons at forward rapidity ($2.5 < y < 4$) using 5.02\TeV \PbPb collisions~\cite{Acharya:2019hlv}.  

This paper reports on the measurement of \vtwo coefficients using the scalar product method~\cite{PhysRevC.87.044907, SPRef:2008} for \PgUa and \PgUb mesons at midrapidity ($\abs{y}<2.4$) in \PbPb collisions at $\sqrtsNN = 5.02\TeV$. The data set was collected with the CMS detector in 2018 and corresponds to an integrated luminosity of 1.7\nbinv. The \PgU mesons are reconstructed through their \dimu decay channel. 
The \vtwo values for \PgUa mesons are measured in four transverse momentum (\pt) intervals (0--3, 3--6, 6--10, and 10--50\GeVc) for different centrality classes.
Centrality is defined as the fraction of the total inelastic nucleus-nucleus cross section with 0\% representing the largest overlap of the two colliding Pb nuclei.
For this analysis, events are classified in three centrality intervals (10--30\%, 30--50\%, and 50--90\%), with 10--30\% corresponding to the most central events.  More central (0--10\%) and more peripheral (90--100\%) collisions are excluded because of small eccentricity in the initial geometry and low event selection efficiency, respectively.  
Because of their lower production cross section compared to that of \PgUa mesons, the \PgUb mesons are measured in one interval that spans $\abs{y}<2.4$, $0 < \pt < 50 \GeVc$, and centrality 10--90\% .

\section{The CMS Detector}
\label{sec:CMSdet}

The central feature of the CMS apparatus is a superconducting solenoid of 6\unit{m} internal diameter, providing a magnetic field of 3.8\unit{T}. Within the solenoid volume are a silicon pixel and strip tracker, a lead tungstate crystal electromagnetic calorimeter, as well as a brass and scintillator hadron calorimeter, each composed of a barrel and two endcap sections. The Hadronic Forward (HF) calorimeters, located at $ 3 < \abs{\eta} < 5$, extend the pseudorapidity ($\eta$) coverage provided by the barrel and endcap detectors.
The collision centrality is estimated experimentally using the sum of the transverse energy deposited in the HF calorimeters.
Muons are detected in gas-ionization chambers embedded in the steel flux-return yoke outside the solenoid within the pseudorapidity coverage of $\abs{\eta}<2.4$.
 A more detailed description of the CMS detector, together with a definition of the coordinate system used and the relevant kinematic variables, can be found in Ref.~\cite{Chatrchyan:2008zzk}.

\section{Event Selection}
\label{sec:EvtSel}

Events were selected based on several trigger stages.  First, at least two muon candidates reconstructed in the fast hardware-based trigger system are required.  In the high level trigger, Level-2 (L2) muons are identified by requiring fitted tracks in the outer muon spectrometer.  At Level-3 (L3), muons are reconstructed using full tracks from the L2 muon tracks and the inner tracker information, with a minimum of ten high-quality hits in the inner tracker~\cite{Khachatryan:2017trig}.  Finally, events are required to contain at least one L2 muon and another L3 muon, with their invariant mass being greater than 7\GeVcc.
To reject beam-related background processes (beam scraping events and beam-gas collisions), events are required to pass the following conditions: (a) that at least one reconstructed primary vertex exists, (b) that the width of silicon pixel detector clusters is compatible with the vertex position~\cite{CMS:2010za}, and (c) that at least two towers in each of the two HF detectors located on opposite sides of CMS detect energy above 4\GeV.
The primary vertex, which is reconstructed based on two tracks or more, is required to be located within 15\unit{cm} from the central point of the detector along the beam axis.

For high-quality muon identification, a set of offline selection criteria is applied.  This selection requires that the reconstructed muon is matched to a track reconstructed in the inner tracking detector, which has at least six silicon tracker hits, with at least one of these hits occurring in the inner silicon pixel detector array.  In addition, the distance of closest approach of a muon track from the nearest primary vertex must be less than 0.3 and 20\unit{cm} in the transverse and the longitudinal directions, respectively. To assure high efficiency, the individual muons were required to have $\pt^{\mu} > 3.5\GeVc$, with $\abs{\eta^{\mu}} < 2.4$. 
Pairs of oppositely charged muons are fitted with a common vertex constraint and used in this analysis if the confidence level of the fit is greater than 1\%.

\section{Analysis}
\label{sec:Ana}
 
The \PgU meson signals and combinatorial backgrounds for selected \dimu pairs are separated using fits to the invariant mass ($m_\text{inv}$) distribution.  The range of the studied mass spectrum is 8--14\GeVcc, which covers both \PgUa and \PgUb resonance peaks, as well as the \PgUc mesons.

Acceptance and efficiency correction factors are used as event-by-event weights when the invariant mass distribution and the  average \vtwo values are computed.
Acceptance is defined as the probability for both the decay muons originated from an \PgU meson to be within $\pt^{\mu} > 3.5\GeVc$, $\abs{\eta^{\mu}} < 2.4$.  It is determined as a function of \pt using \PYTHIA 8.212~\cite{Sjostrand:2014zea} Monte Carlo (MC) generator with tune CP5~\cite{cp5tune}, where \PgU mesons from $^{3}S_{1}$, $^{3}P_{J}$, $^{3}D_{J}$ bottomonium states form via the color-singlet~\cite{Baier:1983va,Gastmans:1986qv} and color-octet~\cite{Cho:1995ce} mechanisms. The MC events are then reweighted to match the \pt spectra to the PbPb data.  
The efficiencies for reconstruction, muon identification, and trigger selection are calculated by a full simulation of the  CMS detector using \GEANTfour~\cite{Agostinelli:2002hh}. Each \PgU event is embedded in a \HYDJET 1.9~\cite{Lokhtin:2008xi} PbPb event to reproduce the hydrodynamic background environment.  The average acceptance and efficiency factors are 53.7\% (56.7\%) and 56.9\% (60.0\%) for \PgUa (\PgUb), respectively.
A systematic uncertainty is assigned based on the difference found between the MC results and experimental data.
The MC and experimental data are compared using the tag-and-probe method with single muons obtained from \PJGy meson decays~\cite{Khachatryan:2011mwz, Chatrchyan:2012xi}.  The efficiency ratio of data over simulation is applied as a weight for the derivation of the final \dimu efficiency.  

The $v_{2}$ values of \PgU candidates are determined using the scalar product (SP) method, in a similar way as the one employed to determine the elliptic flow of 
prompt $D^{0}$ mesons in PbPb collisions~\cite{Sirunyan:2017plt}.  
Using this method, the particle asymmetries associated with different subevents are characterized by $Q$-vectors that are determined for different $\eta$ ranges using either tracker or HF calorimeter data~\cite{PhysRevC.87.044907}.  The $Q$-vectors are given by $Q_{2} = \sum_{k=1}^{M} \omega_{k} \re^{2i\phi_{k}}$, where the sum is over the multiplicity of particles for the tracker or the number of towers for the HF calorimeters, $\phi$ is the azimuthal angle of the particle or the tower, and $\omega$ is the weight given by the \pt of a particle in the tracker or the transverse energy deposited in a tower.  A three subevent method~\cite{Sirunyan:2017plt,PhysRevC.87.044907} is used, employing the tracker, with $\abs{\eta} < 0.75$, and the two HF calorimeters that cover the ranges  $3 < \eta < 5$ (HF+) and $-5 < \eta < -3$ (HF--).  
 The $Q$-vector of an \PgU candidate is defined as $Q_{2,\PgU} = \re^{i2\phi}$ where $\phi$ is the azimuthal angle of the candidate.  
The \vtwo coefficient is then given by
\begin{linenomath*}
\begin{equation}
\label{eq:qsp}
{v_2}\left\{ {{\mathrm{SP}}} \right\} \equiv {\frac{\left\langle {{Q_{2,\PgU}}Q_{2A}^*} \right\rangle }  {\sqrt {{\frac{\left\langle {Q_{2A}^{}Q_{2B}^*} \right\rangle \left\langle {Q_{2A}^{}Q_{2C}^*} \right\rangle } {\left\langle {Q_{2B}^{}Q_{2C}^*} \right\rangle }}} }}.
\end{equation}
\end{linenomath*}
Here, $Q_{2A}$ and $Q_{2B}$ correspond to subevents based on the HF detectors, and $Q_{2C}$ is based on the tracker.  The subscripts $A$ and $B$ refer to either HF+ or HF--, depending on the rapidity of the \PgU candidate.
In order to avoid  autocorrelations and to reduce nonflow effects, the $\eta$ gap between \PgU candidates and the subevent detector is required to be at least 3 units.  For this reason, HF+ is selected for A (B) when the \PgU candidate is produced at negative (positive) rapidity.

\begin{figure}[htb]
 \centering
   \includegraphics[width=0.49\textwidth]{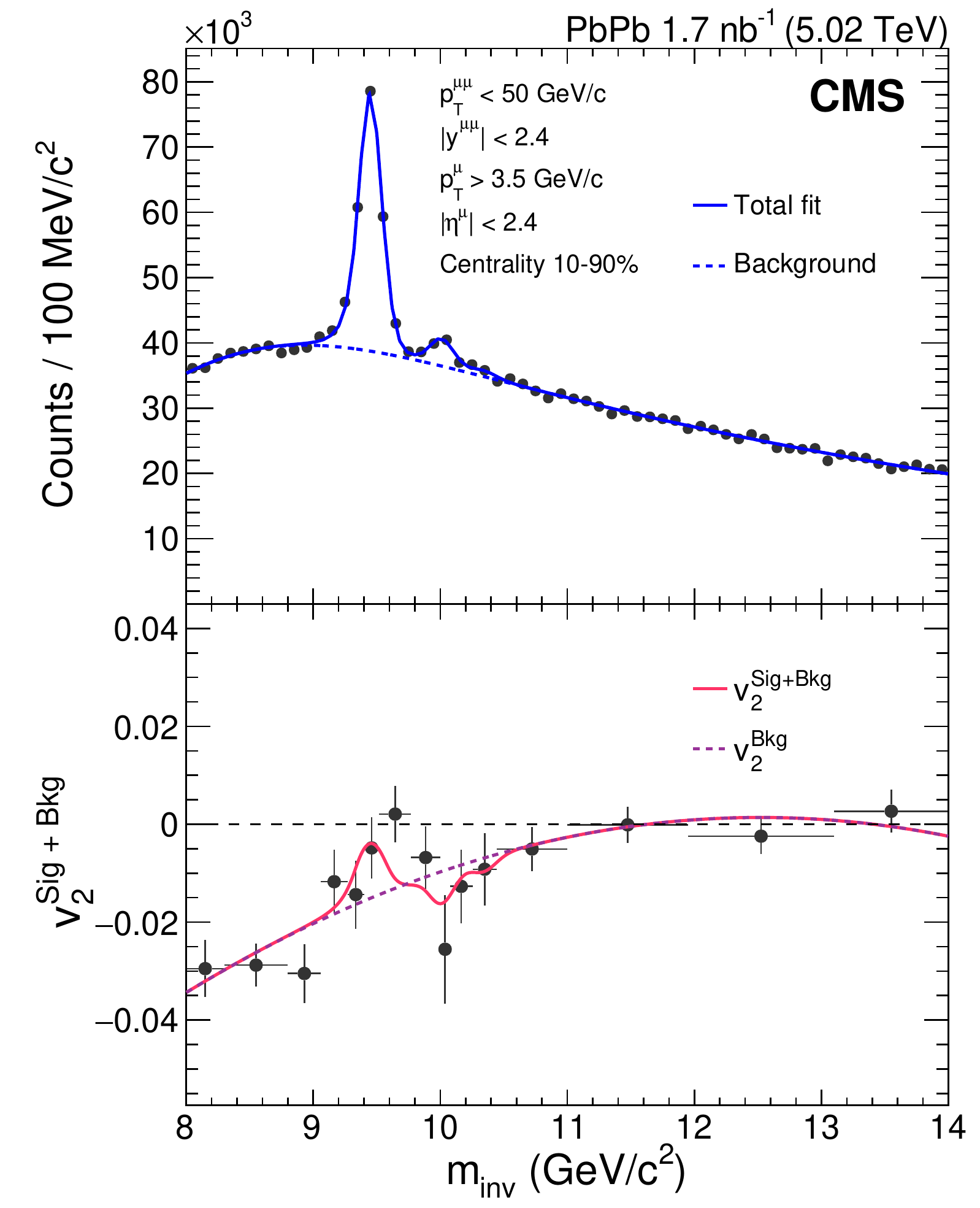}
   \caption{Simultaneous fit of the \dimu invariant mass spectrum and the $\vtwo^{\mathrm{Sig+Bkg}}$ distribution, as defined in Eq.~(\ref{eq:v2mass}), for $\pt < 50\GeVc$ and with centrality 10--90\%. The solid (signal + background) and dashed (background only) blue lines show the result of the mass fit, and the solid and dashed red lines show the corresponding results for the binned $\chi^2$ fit to the \vtwo distribution. } 
  \label{fig:simulFit}
\end{figure}

The $\left\langle \right\rangle$ brackets denote averages over all \PgU candidate events, taking the
real part of the $Q$-vector products. The denominator in Eq.~(\ref{eq:qsp}) is a resolution correction that depends on the particle multiplicity in the HF detector referenced by the A subscript~\cite{PhysRevC.87.044907}. The lower panel of Fig.~\ref{fig:simulFit} shows an example of the resulting \vtwo distribution as a function of the \PgU candidate invariant mass.
The separation of signal and background is done in two steps.  First, the yields of \PgU signals are extracted from the invariant mass distribution without using the \vtwo information.  The purpose of this step is to find the best parameters for the probability distribution describing the  \PgU mass peaks. Unbinned log-likelihood fits to the mass distribution are used to obtain the best fit to the data. 
The analysis is done using a single rapidity range with $\abs{y}<2.4$ and, thus, uses \PgU candidates obtained in both the barrel and endcap regions of CMS. For this reason, 
the signal mass distribution (Sig) for each \PgU meson is formed by the sum of two Crystal Ball (CB) functions~\cite{CrystalBallRef}, to  account for the different mass resolutions of the two regions.    For both CB functions, the mass and the radiative tail parameters are constrained to be equal since these are not sensitive to the detector resolution. The  \PgUa mass is taken as a free parameter of the fit, allowing for a possible scaling error in the momentum calibration for the reconstructed tracks.  The mean ($m_{0}$) and width ($\sigma$) parameters for the excited states (\PgUb and \PgUc) are found by scaling the \PgUa values by the ratio of published masses~\cite{PDG2018}.  Other parameters for the excited states are constrained to be identical to those of the \PgUa mesons.  The normalization parameters \NSigone, \NSigtwo and \NSigthree~are free parameters of the fit.  More details about the parameters for the invariant mass fit can be found in Ref.~\cite{Sirunyan:2018nsz}.

The background mass spectrum (Bkg) is modeled by a function obtained by multiplying a real-valued error function and an exponential function.  The error function is used to reproduce a kinematic enhancement found at low mass, resulting from the single muon \pt threshold.  The exponential component is motivated by the exponentially falling structure of the combinatorial background in the high mass region.  There are four parameters characterizing the shape of background, the $\mu$ and $\sigma$ parameters of the error function, the decay length of the exponential function, and an overall normalization parameter.  

In the second step, the invariant mass and \vtwo distributions are fitted simultaneously using binned $\chi^2$ fits.
The signal function parameters are fixed to the values obtained from the previous step, although the normalizations \NSigone, \NSigtwo, and \NSigthree\ and the background function parameters are left free.
The \vtwo dependence on $m_\text{inv}$ is taken as
\begin{linenomath}
\label{vnvsmassequation}
\ifthenelse{\boolean{cms@external}}
{
\begin{multline}
  \label{eq:v2mass}
  v_{2}^{\mathrm{Sig+Bkg}}(m_\text{inv}) = \alpha_{1}(m_\text{inv})v_{2}^{\PgUa} + \alpha_{2}(m_\text{inv})v_{2}^{\PgUb}  \\
+ \alpha_{3}(m_\text{inv})v_{2}^{\PgUc} + [ 1-\alpha_{1}(m_\text{inv})  \\
 - \alpha_{2}(m_\text{inv}) - \alpha_{3}(m_\text{inv}) ] v_{2}^{Bkg}(m_\text{inv}),  
\end{multline}
}
{
\begin{equation}
  \label{eq:v2mass}
  \begin{aligned}
v_{2}^{\mathrm{Sig+Bkg}}(m_\text{inv}) = \alpha_{1}(m_\text{inv})v_{2}^{\PgUa} + \alpha_{2}(m_\text{inv})v_{2}^{\PgUb} + \alpha_{3}(m_\text{inv})v_{2}^{\PgUc} \\
    + [ 1-\alpha_{1}(m_\text{inv}) - \alpha_{2}(m_\text{inv}) - \alpha_{3}(m_\text{inv}) ] v_{2}^{Bkg}(m_\text{inv}),
  \end{aligned}
\end{equation}
}
\end{linenomath}
where, $v_{2}^{\PgU(iS)}$ ($i= 1, 2, 3$) is the \vtwo value for the \PGUP{\textit{i}S} mesons and is assumed to be independent of $m_\text{inv}$. The $\alpha_{i}(m_\text{inv})$ coefficients are the fractions that the \PGUP{\textit{i}S} states occupy as a function of invariant mass, as determined from the fit. They are defined as 
\begin{linenomath}
\label{vnvsmasseq2}
\ifthenelse{\boolean{cms@external}}
{
\begin{multline}
\alpha_{i}(m_\text{inv}) = \mathrm{Sig}_{\PGUP{\textit{i}S}}(m_\text{inv}) / [ \mathrm{Sig}_{\PgUa}(m_\text{inv})  \\
  + \mathrm{Sig}_{\PgUb}(m_\text{inv}) + \mathrm{Sig}_{\PgUc}(m_\text{inv}) + \text{Bkg}(m_\text{inv}) ].  \\ 
\end{multline}
}
{
\begin{equation}
\alpha_{i}(m_\text{inv}) = \mathrm{Sig}_{\PGUP{\textit{i}S}}(m_\text{inv}) / [ \mathrm{Sig}_{\PgUa}(m_\text{inv}) + \mathrm{Sig}_{\PgUb}(m_\text{inv}) + \mathrm{Sig}_{\PgUc}(m_\text{inv}) + \text{Bkg}(m_\text{inv}) ].
\end{equation}
}
\end{linenomath}
The background \vtwo value, $\vtwo^{{\text{Bkg}}}(m_\text{inv})$, is modeled as a second-order polynomial function of the invariant mass.
Figure~\ref{fig:simulFit} shows an example  of a simultaneous fit to the mass (upper panel) and \vtwo distributions (lower panel).

\section{Systematic Uncertainties}
\label{sec:SysUnc}

The systematic uncertainties in this analysis come from various sources, including the modeling of signal and background probability functions, as well as acceptance and efficiency corrections (muon identification and trigger). The uncertainties for the various sources are studied as functions of \pt and centrality. However, within the limits of the evaluation, no clear \pt or centrality trend is found. For the uncertainty associated with the signal, an alternative function is formed by adding one CB and one Gaussian function having equal mass parameters.  The differences in the \vtwo values from the nominal one are taken as the uncertainties. The difference is typically less than 0.0025 for the \PgUa mesons and is equal to 0.014 for the \PgUb mesons. The uncertainties resulting from holding the final fit parameters constant are studied by releasing each parameter, one by one, and redoing the fit with the additional free parameter.  The largest deviation of the resulting \vtwo values from the nominal result is taken as the uncertainty. The difference is found typically in the range 0.0006 to 0.005 for the \PgUa mesons and is 0.0093 for the \PgUb mesons. To determine the uncertainty related to the background modeling, the functions for both the invariant mass distribution and $\vtwo^{{\text{Bkg}}}(m_\text{inv})$ are varied. For the mass distribution, a fourth-order polynomial is used as an alternative functional behavior. For the $\vtwo^{{\text{Bkg}}}(m_\text{inv})$ function, two alternative functions are introduced instead of the nominal second-order polynomial, one as the third-order polynomial and the other as an exponential function.
The uncertainty associated with the choice of the $\vtwo^{{\text{Bkg}}}(m_\text{inv})$ function is assigned as the maximum difference found for the two alternative functions used to model the background. 
The uncertainty values in the choice of the background modeling functions for the mass and \vtwo distributions are typically in the range 0.001--0.009 for the \PgUa mesons, and are found to be 0.012 and 0.030, respectively for the \PgUb mesons.

The uncertainty resulting from the acceptance and efficiency corrections is evaluated by comparing the results with and without weighting the simulated \pt spectra. By virtue of the cylindrical symmetry of the detector, the effect of these corrections is minor. An additional systematic uncertainty for the efficiency is assigned based on the tag-and-probe method. The difference of the \vtwo values with and without applying the tag-and-probe correction is assigned as the uncertainty and is found to be in the range 0.002--0.012. Finally, uncertainties resulting from the hadronic event selection are considered.  The collision event filter is varied, thus allowing for a migration of \PgUa and \PgUb mesons across centrality boundaries. The effect on the measured \vtwo value is taken as a systematic uncertainty. For \PgUa the differences are typically below 0.0008 and for \PgUb the uncertainty is 0.0067. Individual systematic uncertainties are expected to be uncorrelated and are therefore added in quadrature for the final assigned values.
The absolute systematic uncertainty in the \vtwo values for the \PgUa meson range in most intervals from 0.002 to 0.015, which is smaller than the statistical uncertainty. The largest uncertainty corresponds to the most peripheral and lowest \pt interval measured, which is dominated by the uncertainty related to the $v_{2}^{{\text{Bkg}}}(m_\text{inv})$ modeling and hadronic event selection. The uncertainty for the \PgUb \vtwo value, which is obtained for the integrated interval only, is 0.037.  

\section{Results}
\label{sec:Res}

\begin{figure}[htb!p]
 \centering
      \includegraphics[width=0.49\textwidth]{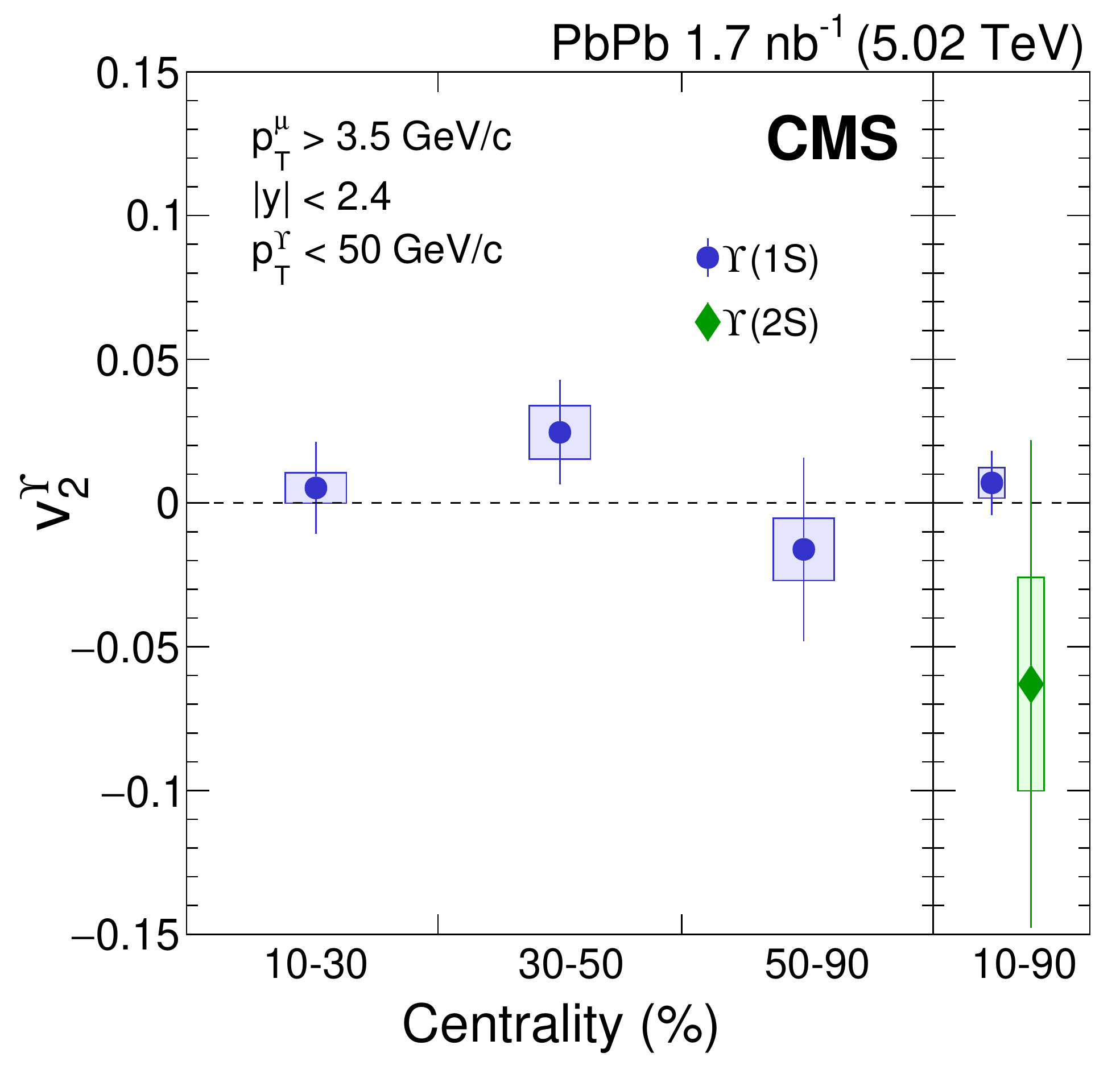}
      \includegraphics[width=0.49\textwidth]{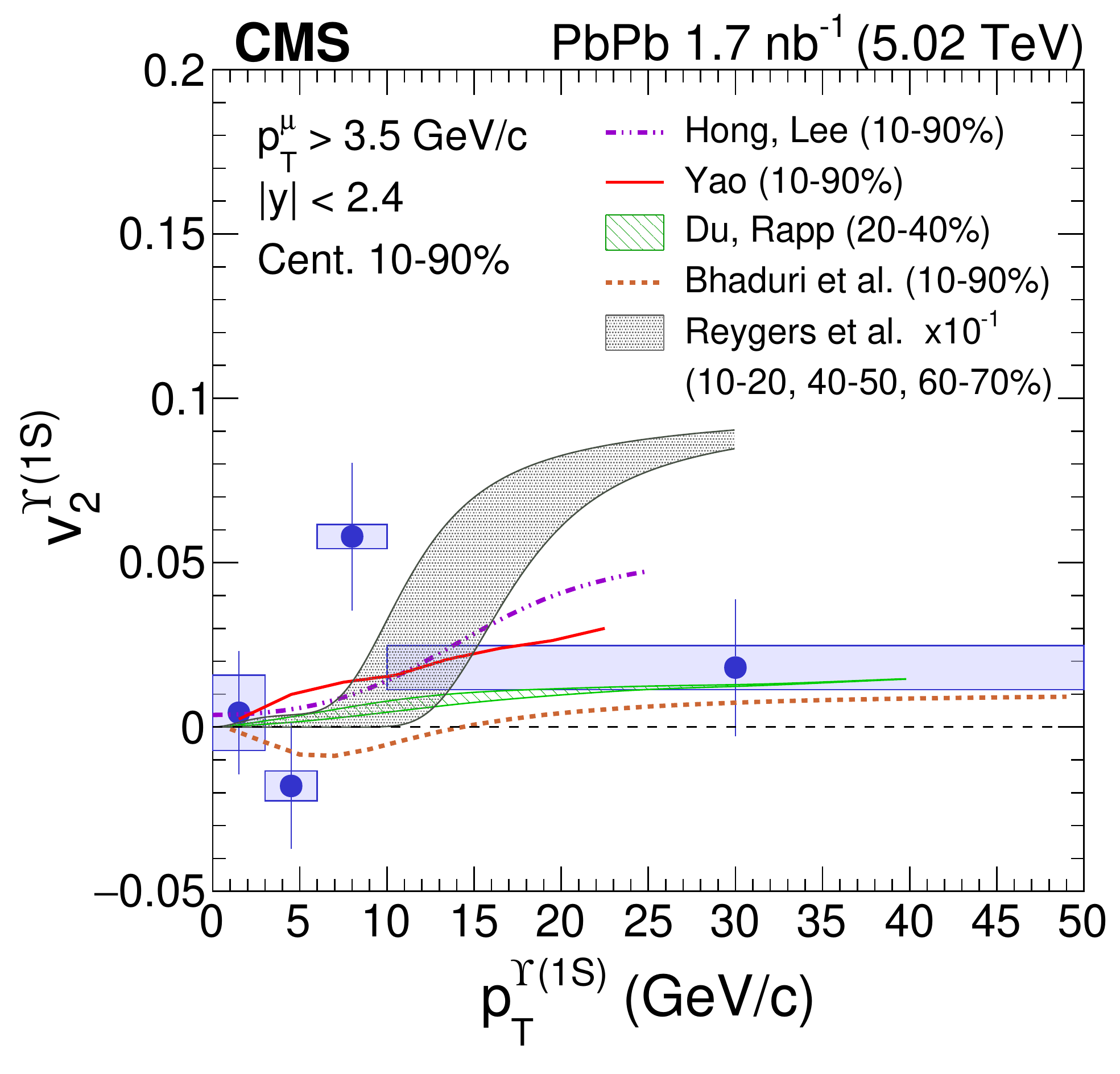}
   \caption{(\cmsLeft panel) \pt integrated \vtwo values for \PgUa mesons measured in four centrality classes and for the \PgUb meson in the 10--90\% centrality range. (\cmsRight panel) \vtwo of \PgUa mesons as a function of \pt in the 10--90\% centrality range compared with model calculations from Du and Rapp~\cite{Du:2017qkv}, Yao~\cite{Yao:2020CoupledQuark,YaoQuarkInsideQGP}, Hong and Lee~\cite{Hong:2019ade, Ke:2018tsh}, Bhaduri et al.~\cite{Bhaduri2020anisotropic}, and Reygers et al.~\cite{Reygers:2019aul}.  All results are for the rapidity range $\abs{y}<2.4$.  The vertical bars denote statistical uncertainties, and the rectangular boxes show the total systematic uncertainties. }
   \label{fig:resultInt}
\end{figure}

The \pt integrated results are shown in Fig.~\ref{fig:resultInt} (\cmsLeft panel) for three centrality intervals.
The \PgUa \vtwo values are consistent with zero within the statistical uncertainties.
The rightmost points in this figure are the average \vtwo values in the 10--90\% centrality interval. 
They are determined to be $0.007\pm 0.011\stat\pm 0.005\syst$ for \PgUa mesons and $-0.063\pm 0.085\stat\pm 0.037\syst$ for \PgUb mesons.

\begin{figure*}
 \centering
   \includegraphics[width=0.95\textwidth]{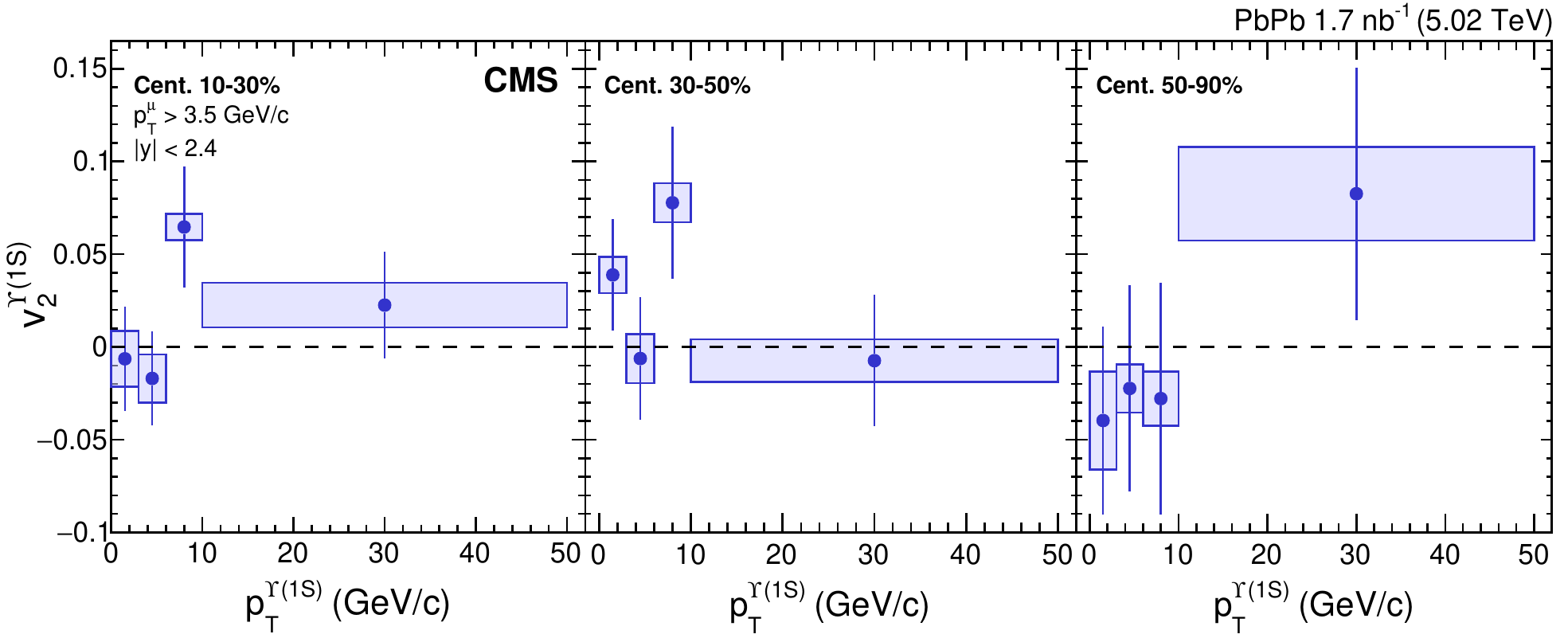}
   \caption{The \vtwo coefficients for \PgUa mesons as a function of \pt in three centrality classes: 10--30\% (left panel), 30--50\% (middle panel) and 50--90\% (right panel). 
   The rapidity range is $\abs{y}<2.4$.  The vertical lines indicate the  statistical uncertainties and the rectangular boxes show the total systematic uncertainties. 
   }
   \label{fig:resultCentBin}
\end{figure*}

\begin{figure}[htbp]
 \centering
   \includegraphics[width=0.49\textwidth]{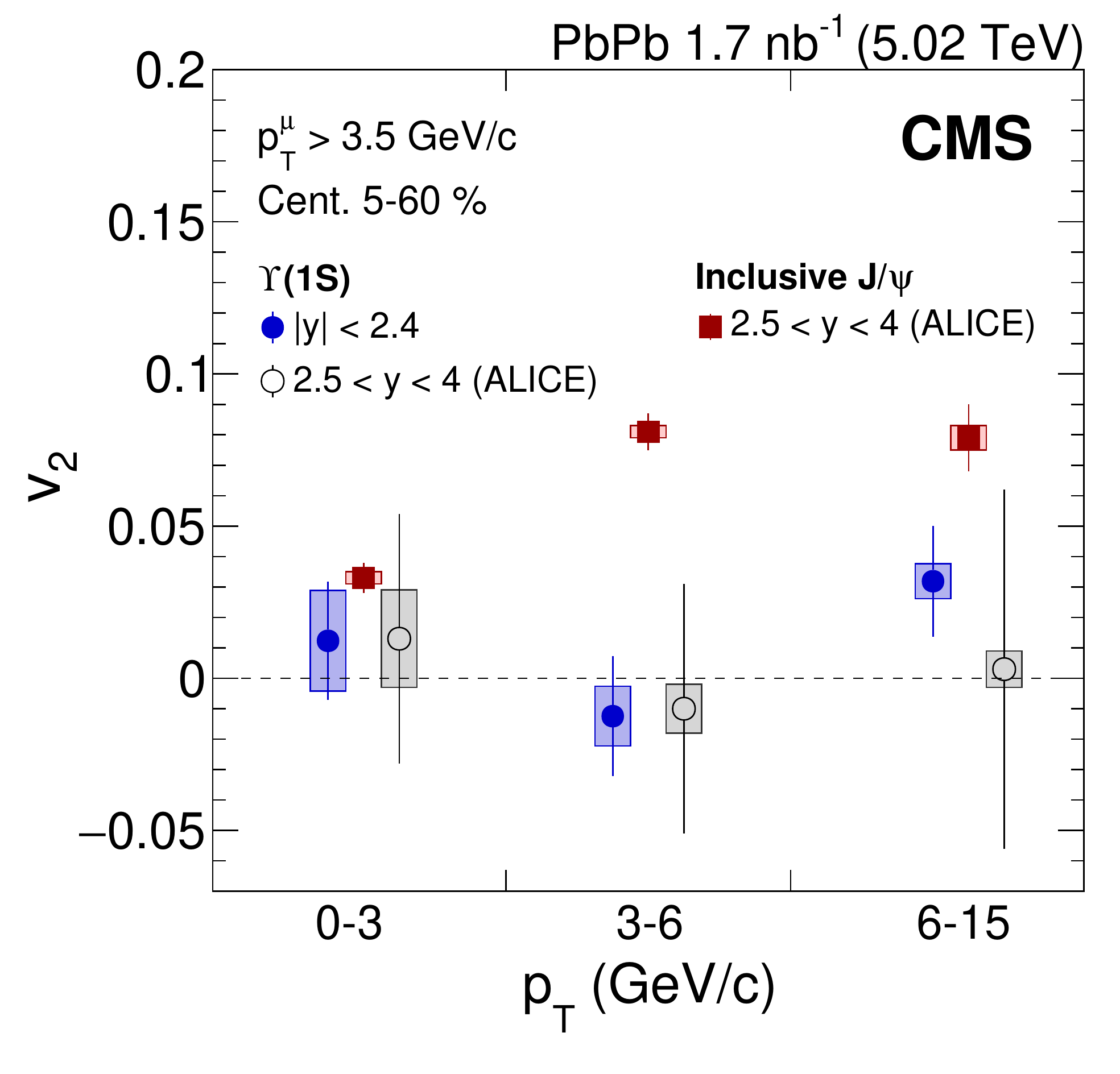}
   \caption{The \vtwo for \PgUa mesons as a function of \pt in the rapidity range $\abs{y}<2.4$ compared with the ALICE results for \PgUa (open circles) and \PJGy (full squares) mesons measured in $2.5<y<4$~\cite{Acharya:2019hlv}. All results are obtained in the range $0 < \pt < 15\GeVc$ and centrality interval 5--60\%. The vertical bars denote statistical uncertainties, and the rectangular boxes show the total systematic uncertainties.} 
   \label{fig:result_CompALICE}
\end{figure}

In Fig.~\ref{fig:resultInt} (\cmsRight panel), the \pt dependence of \PgUa meson \vtwo values is measured for the 10--90\% centrality interval. The \vtwo values are consistent with zero in the measured \pt range, except for the $6 < \pt < 10\GeVc$ interval that shows a $2.6~\sigma$ deviation from zero.
The result is also compared with theoretical predictions from five different approaches. 
The green shaded area (Du, Rapp~\cite{Du:2017qkv}) is a calculation using the kinetic rate equation framework for the time evolution of bottom quarks. In this model, the in-medium effect is applied with a lattice QCD based equation of state.  Note that it is calculated for the centrality interval 20--40\%.  The red line (Yao et al.~\cite{Yao:2020CoupledQuark,YaoQuarkInsideQGP}) is computed from a real-time simulation of heavy quarks using coupled Boltzmann transport equations.  The dissociation and recombination rates are calculated from potential non-relativistic QCD for this model. The dashed violet curve (Hong, Lee~\cite{Hong:2019ade, Ke:2018tsh}) is another kinetic model prediction in which the heavy quark collisional terms are obtained from Bethe-Salpeter amplitude and hard thermal loop resummation.  A diffusion constant $D(2 \pi T)~=~6$, where T is the temperature of the medium, was used.  With this parameter, the nuclear modification factor, which is the ratio of the quarkonium yield in a nucleus-nucleus collision to the corresponding yield in a proton-proton (pp) collision scaled by the number of binary collisions, is consistent with the CMS result in Ref.~\cite{Sirunyan:2018nsz}.
The dashed brown line (Bhaduri et al.~\cite{Bhaduri2020anisotropic}) shows the result of 3+1d quasiparticle anisotropic hydrodynamics framework.  The initial conditions for temperature and shear viscosity to entropy density ratio are tuned to the identified hadron spectra and flow harmonics in 5.02\TeV PbPb data~\cite{BhaduriIdentifiedHadronALICE:2020}.  The gray shaded area (Reygers et al.~\cite{Reygers:2019aul}) is another hydrodynamics calculation based on the blast-wave fits using the data of lighter particles, which was scaled down by a factor of 1/10.  The  \vtwo values are computed in three centrality intervals (10--20\%, 40--50\%, 60--70\%) and the gray envelope is constructed to minimally cover all results. 
Despite their slightly different \pt dependence, all models in the figure derive small \vtwo values for $\pt < 10 \GeVc$, which is consistent with the experimental result.  For the highest \pt range, however, the kinetic model (Hong, Lee) and the blast-wave model begin to deviate from the observed behavior.
In particular, the blast-wave model predicts a \vtwo value much larger than found.
In this model, the elliptic flow of \PgUa is expected to sharply rise from $\pt = 9$ \GeVc, following the mass ordering of \vtwo(\pt) by lighter particles used for the fit~\cite{Reygers:2019aul}.

The \pt differential results for \PgUa mesons in each centrality interval are shown in Fig.~\ref{fig:resultCentBin} and the \vtwo values are found to be consistent with zero within uncertainties.
A similar result has been obtained by the ALICE Collaboration~\cite{Acharya:2019hlv}, where the measurement was done in a complementary rapidity region ($2.5<y<4$), as shown in Fig.~\ref{fig:result_CompALICE}. The CMS data points in this figure were analyzed using the same \pt range as the ALICE result to allow for a comparison where only the rapidity ranges of the two measurements differ.
The ALICE Collaboration also measured inclusive \PJGy \vtwo values (also shown in Fig.~\ref{fig:result_CompALICE}) with the same kinematic conditions as their \PgUa results and there find a significant and finite \vtwo value.
Together, the CMS and ALICE results indicate that the geometry of the medium has little influence on the \PgUa yields~\cite{Du:2017qkv, bhaduri2018anisotropic} and that recombination is not a dominant process in the production of this meson. The results also indicate that the path-length dependence of  \PgUa suppression is small.

The current results support the assumption of a high melting temperature of the \PgUa meson.    Consequently, the dissociation can only take place in the earliest stages of the collision.  This makes the \PgUa meson less  sensitive to the path length than other quarkonia states.

\section{Summary}
\label{sec:Sum}

The \vtwo coefficients for \PgUa and \PgUb mesons are measured in PbPb collisions at a nucleon-nucleon center-of-mass energy of 5.02\TeV.  Results are reported for the rapidity range $\abs{y}<2.4$, in the transverse momentum interval $0 < \pt < 50\GeVc$, and in three centrality classes of 10--30\%, 30--50\%, and 50--90\% for the \PgUa meson, while the centrality interval 10--90\% is used for the \PgUb meson.  The \vtwo values are observed to be compatible with zero for the \PgUa meson in the measured kinematic and centrality intervals.
The observation contrasts with the positive \vtwo values reported for \PJGy mesons, suggesting different medium effects for charmonia and bottomonia.
This confirms, with higher statistical precision and with an extended \pt range, the result from the ALICE experiment measured in a complementary rapidity region.
The measured values of \vtwo for $\pt < 10 \GeVc$ are consistent with the theoretical predictions based on the kinetic model, the hydrodynamics framework, and the blast-wave model.  For the high \pt region, the  \PgUa result provides an important constraint for models.
The \vtwo value found for \PgUb mesons, which is being reported for the first time, is also consistent with zero.
As there are expected to be differences in the various processes through which the QGP affects \PgUa and \PgUb mesons, these measurements provide new inputs for the study of bottomonia production in heavy ion collisions.

\begin{acknowledgments}
  We congratulate our colleagues in the CERN accelerator departments for the excellent performance of the LHC and thank the technical and administrative staffs at CERN and at other CMS institutes for their contributions to the success of the CMS effort. In addition, we gratefully acknowledge the computing centres and personnel of the Worldwide LHC Computing Grid for delivering so effectively the computing infrastructure essential to our analyses. Finally, we acknowledge the enduring support for the construction and operation of the LHC and the CMS detector provided by the following funding agencies: BMBWF and FWF (Austria); FNRS and FWO (Belgium); CNPq, CAPES, FAPERJ, FAPERGS, and FAPESP (Brazil); MES (Bulgaria); CERN; CAS, MoST, and NSFC (China); COLCIENCIAS (Colombia); MSES and CSF (Croatia); RPF (Cyprus); SENESCYT (Ecuador); MoER, ERC IUT, PUT and ERDF (Estonia); Academy of Finland, MEC, and HIP (Finland); CEA and CNRS/IN2P3 (France); BMBF, DFG, and HGF (Germany); GSRT (Greece); NKFIA (Hungary); DAE and DST (India); IPM (Iran); SFI (Ireland); INFN (Italy); MSIP and NRF (Republic of Korea); MES (Latvia); LAS (Lithuania); MOE and UM (Malaysia); BUAP, CINVESTAV, CONACYT, LNS, SEP, and UASLP-FAI (Mexico); MOS (Montenegro); MBIE (New Zealand); PAEC (Pakistan); MSHE and NSC (Poland); FCT (Portugal); JINR (Dubna); MON, RosAtom, RAS, RFBR, and NRC KI (Russia); MESTD (Serbia); SEIDI, CPAN, PCTI, and FEDER (Spain); MOSTR (Sri Lanka); Swiss Funding Agencies (Switzerland); MST (Taipei); ThEPCenter, IPST, STAR, and NSTDA (Thailand); TUBITAK and TAEK (Turkey); NASU (Ukraine); STFC (United Kingdom); DOE and NSF (USA). 
  
  \hyphenation{Rachada-pisek} Individuals have received support from the Marie-Curie programme and the European Research Council and Horizon 2020 Grant, contract Nos.\ 675440, 752730, and 765710 (European Union); the Leventis Foundation; the A.P.\ Sloan Foundation; the Alexander von Humboldt Foundation; the Belgian Federal Science Policy Office; the Fonds pour la Formation \`a la Recherche dans l'Industrie et dans l'Agriculture (FRIA-Belgium); the Agentschap voor Innovatie door Wetenschap en Technologie (IWT-Belgium); the F.R.S.-FNRS and FWO (Belgium) under the ``Excellence of Science -- EOS" -- be.h project n.\ 30820817; the Beijing Municipal Science \& Technology Commission, No. Z191100007219010; the Ministry of Education, Youth and Sports (MEYS) of the Czech Republic; the Deutsche Forschungsgemeinschaft (DFG) under Germany's Excellence Strategy -- EXC 2121 ``Quantum Universe" -- 390833306; the Lend\"ulet (``Momentum") Programme and the J\'anos Bolyai Research Scholarship of the Hungarian Academy of Sciences, the New National Excellence Program \'UNKP, the NKFIA research grants 123842, 123959, 124845, 124850, 125105, 128713, 128786, and 129058 (Hungary); the Council of Science and Industrial Research, India; the HOMING PLUS programme of the Foundation for Polish Science, cofinanced from European Union, Regional Development Fund, the Mobility Plus programme of the Ministry of Science and Higher Education, the National Science Center (Poland), contracts Harmonia 2014/14/M/ST2/00428, Opus 2014/13/B/ST2/02543, 2014/15/B/ST2/03998, and 2015/19/B/ST2/02861, Sonata-bis 2012/07/E/ST2/01406; the National Priorities Research Program by Qatar National Research Fund; the Ministry of Science and Higher Education, project no. 02.a03.21.0005 (Russia); the Tomsk Polytechnic University Competitiveness Enhancement Program and ``Nauka" Project FSWW-2020-0008 (Russia); the Programa Estatal de Fomento de la Investigaci{\'o}n Cient{\'i}fica y T{\'e}cnica de Excelencia Mar\'{\i}a de Maeztu, grant MDM-2015-0509 and the Programa Severo Ochoa del Principado de Asturias; the Thalis and Aristeia programmes cofinanced by EU-ESF and the Greek NSRF; the Rachadapisek Sompot Fund for Postdoctoral Fellowship, Chulalongkorn University and the Chulalongkorn Academic into Its 2nd Century Project Advancement Project (Thailand); the Kavli Foundation; the Nvidia Corporation; the SuperMicro Corporation; the Welch Foundation, contract C-1845; and the Weston Havens Foundation (USA). 
\end{acknowledgments}

\bibliography{auto_generated} 

\cleardoublepage \appendix\section{The CMS Collaboration \label{app:collab}}\begin{sloppypar}\hyphenpenalty=5000\widowpenalty=500\clubpenalty=5000\vskip\cmsinstskip
\textbf{Yerevan Physics Institute, Yerevan, Armenia}\\*[0pt]
A.M.~Sirunyan$^{\textrm{\dag}}$, A.~Tumasyan
\vskip\cmsinstskip
\textbf{Institut f\"{u}r Hochenergiephysik, Wien, Austria}\\*[0pt]
W.~Adam, F.~Ambrogi, T.~Bergauer, M.~Dragicevic, J.~Er\"{o}, A.~Escalante~Del~Valle, M.~Flechl, R.~Fr\"{u}hwirth\cmsAuthorMark{1}, M.~Jeitler\cmsAuthorMark{1}, N.~Krammer, I.~Kr\"{a}tschmer, D.~Liko, T.~Madlener, I.~Mikulec, N.~Rad, J.~Schieck\cmsAuthorMark{1}, R.~Sch\"{o}fbeck, M.~Spanring, W.~Waltenberger, C.-E.~Wulz\cmsAuthorMark{1}, M.~Zarucki
% ...
\vskip\cmsinstskip
\dag: Deceased\\
1:  Also at Vienna University of Technology, Vienna, Austria\\
% ...
\end{sloppypar}
%%% END EDITABLE REGION %%%
% skeleton_end
\end{document}